\documentstyle[12pt]{article}

\newcommand{\Rocek}{Ro\v cek}

\newcommand{\half}{\frac{1}{2}}

\newcommand{\th}{\theta}
\newcommand{\bth}{\bar{\theta}}

\newcommand{\tK}{\widetilde{K}}

\newcommand{\Q}{\widetilde{Q}}

\newcommand{\Ld}{\Lambda}
\newcommand{\bLd}{\overline \Lambda}

\newcommand{\ld}{\lambda}
\newcommand{\bld}{\overline \lambda}

\newcommand{\bPhi}{\overline \Phi}
\newcommand{\bphi}{\overline \phi}
\newcommand{\tPhi}{\widetilde{\Phi}}
\newcommand{\tbPhi}{\overline{\widetilde{\Phi}}}

\newcommand{\bD}{\overline D}

\newcommand{\bchi}{\overline{\chi}}
\newcommand{\bareta}{\bar{\eta}}
\newcommand{\jp}{J_{(+)}}
\newcommand{\jm}{J_{(-)}}
\newcommand{\jpm}{J_{(\pm )}}
\newcommand{\na}{\nabla}

\newcommand{\tz}{\frac{\theta_{12}}{z_{12}}}

\newcommand{\cQ}{\cal Q}
\newcommand{\cT}{\cal T}
\newcommand{\cI}{\cal I}
\newcommand{\cA}{\cal A}
\newcommand{\cB}{\cal B}
\newcommand{\tzb}{\frac{\bar{\theta}_{12}}{z_{12}}}
\newcommand{\tzzbb}{\frac{\theta_{12} \bar{\theta}_{12}}{z_{12}^2}}
\newcommand{\tzbb}{\frac{\theta_{12} \bar{\theta}_{12}}{z_{12}}}
\newcommand{\Ht}{\tilde{h}}

\newcommand{\strutje}{\rule[-1.5mm]{0mm}{5mm}}

\newcommand{\extraspace}{\addtolength{\abovedisplayskip}{2mm}
                        \addtolength{\belowdisplayskip}{2mm}
                        \addtolength{\abovedisplayshortskip}{2mm}
                        \addtolength{\belowdisplayshortskip}{2mm}}
\newcommand{\be}{\begin{equation}\extraspace}

\newcommand{\ee}{\end{equation}}

\newcommand{\bea}{\begin{eqnarray}\extraspace}
\newcommand{\beastar}{\begin{eqnarray*}\extraspace}
\newcommand{\eea}{\end{eqnarray}}
\newcommand{\eeastar}{\end{eqnarray*}}
\newcommand{\nonu}{\nonumber \\[2mm]}

\newcommand{\suu}{$SU(2)\times U(1)$}
\newcommand{\ie}{{\it i.e.,\ }}
\newcommand{\np}{Nucl.\ Phys.\ }
\newcommand{\pr}{Phys.\ Rev.\ }
\newcommand{\cmp}{Comm.\ Math.\ Phys.\ }
\newcommand{\pl}{Phys.\ Lett.\ }

\begin{document}
\baselineskip 15pt

\noindent Oct. 1991 \hfill IASSNS-HEP-91/69, ITP-SB-91-49\\
$\strutje$ \hfill  LBL-31325, UCB-PTH-91/50\\

\vspace{4mm}

\begin{center}

{\Large SUPERSPACE $WZW$ MODELS AND BLACK HOLES%
\footnote{Contribution to the {\it Workshop on Superstrings and
 Related Topics}, Trieste, Aug.\ 1991. \\Talk given by K. Schoutens.}%
\footnote{Work supported in part by grant NSF-91-08054}%
\footnote{This work was supported in part by the Director,
Office of Energy Research, Office of High Energy and Nuclear Physics,
Division of High Energy Physics of the U.S. Department of Energy
under Contract DE-AC03-76SF00098 and in part by the National Science
Foundation under grant PHY90-21139.} }\\

\vspace{1cm}

{\large M. \Rocek%
\footnote{On leave from ITP, SUNY at Stony Brook}}\\

\vspace{2mm}

{\it School of Natural Sciences \\
     Institute for Advanced Study \\
     Olden Lane, Princeton NJ 08540} \\

\vspace{5mm}

{\large C. Ahn, K. Schoutens} \\

\vspace{2mm}

{\it Institute for Theoretical Physics \\
     State University of New York at Stony Brook \\
     Stony Brook, NY 11794-3840} \\

\vspace{3mm}

{\large and}\\

\vspace{3mm}

{\large A. Sevrin}\\

\vspace{2mm}

{\it Department of Physics, University of California\\
     and\\
     Theoretical Physics Group, Lawrence Berkeley Laboratory \\
     Berkeley, California 94720} \\

\vspace{4mm}

\end{center}

\baselineskip=18pt

\noindent

We show how to write an off-shell action for the
$SU(2)\times U(1)$ supersymmetric WZW model in terms
of $N=2$ chiral and twisted chiral multiplets.
We discuss the $N=4$ supersymmetry of this model
and exhibit the $N=4$ superconformal current algebra.
Finally, we show that the off-shell formulation makes it possible
to perform a duality transformation, which leads to a supersymmetric
sigma model on a manifold with a black hole type singularity.

\vfill

\newpage


\baselineskip=18pt

\setcounter{equation}{0}

Off-shell formulations of WZW-models are known in $N=0,1$
superspace \cite{wzw,paolo}.  For example, the $N=1$ action is
\be
S= \int d^2z \na _+\na _- \left[ (g_{ij} +b_{ij}) \na _+\phi^i \na _-\phi^j
\right] \, ,
\ee
where $\phi^i$ is a unconstrained scalar superfield that coordinatizes
the group manifold, $g_{ij}$ is the metric, and $b_{ij}$ is the
potential for the parallelizing torsion.  It is known that any even
dimensional group allows for an $N=2$ super Kac-Moody symmetry, and a subset
of these models have an $N=4$ symmetry \cite{spin}.  On dimensional grounds, it
is clear that $N=2,4$ superspace actions are simply functions
of the superfields without any derivatives; hence it is not evident how
one can write $g$ and $b$ terms separately.  For example, if one writes
an action that depends on the most familiar $N=2$ scalar multiplet, a chiral
superfield, then one finds that $g$ is necessarily K\" ahler and $b=0$
\cite{bruno}.  For WZW models, $g$ is {\it never}
K\" ahler and $b \neq 0$, so chiral superfields are not enough.  This is a new
feature of extended supersymmetry: the dynamics is not determined entirely
by the form of the action, but also by the kinematical nature of the
superfields.  A particular example of a variant (twisted) scalar multiplet
was introduced by Gates, Hull, and \Rocek\ \cite{twist} (see also
\cite{russians}).  In a recent paper \cite{physlet},
we showed that the \suu\ super WZW model can be formulated in $N=2,4$
superspace using a usual chiral and a twisted chiral superfield.
We also showed that all other WZW models require more exotic
representations.

In this paper we will first briefly review the results of \cite{physlet}
for the off-shell formulation of the \suu\ super WZW model in $N=2$
superspace. We will then focus on the on-shell current algebra, and,
working in chiral $N=2$ superspace, explicitly show how the $N=2$
superconformal algebra can be extended to $N=4$. Finally, we will
go back to the classical level and perform a duality transformation
which leads to a dual sigma model. The latter has the interpretation of
a black hole solution to two-dimensional string theory.

\vspace{8mm}

In $N=2$ superspace, we work with complex left and right handed spinor
derivatives $D_\pm$ satisfying the algebra
\be
\{D_\pm , \overline D_\pm \} = \partial _{\pm \pm} \, ,
\ee
all other anticommutators vanish.  Here $\partial _{++} = \partial _z$, etc.
Chiral superfields obey:
\be
\overline D_\pm \Phi =0 \, , \qquad D_\pm \overline \Phi = 0 \, .
\ee
In contrast, twisted chiral superfields obey \cite{twist}
\be
\overline D_+ \Ld = 0 \, , \; D_- \Ld = 0 \, , \; D_+ \overline \Ld = 0
\, , \; \overline D_- \overline \Ld =0 \, .
\ee
Both superfields can be reduced to $N=1$ superfields as follows:
We define real $N=1$ spinor
derivatives $\na _\pm = D_\pm + \overline D_\pm$ and ``extra''
supersymmetry generators $\widetilde Q _\pm = i (\overline D_\pm - D_\pm )$.
The resulting $N=1$ superfields $\phi , \ld$
are unconstrained scalars with the following
transformations under the extra supersymmetry:
\bea
\Q _\pm \phi = -i \na _\pm \phi \, , \qquad
&& \qquad \Q _\pm \bphi = +i \na _\pm
\bphi  \, ,
\nonu
\Q _+ \ld = -i \na _+ \ld \, , \; \Q _- \ld = +i \na _- \ld \, , && \,
\Q _+ \bld = +i \na _+ \bld \, , \; \Q _- \bld = -i \na _- \bld \,.
\label{eq:susy}
\eea
However, it is known that extra supersymmetries can be written in $N=1$
superspace as \cite{twist}
\be
\Q _\pm \phi ^i = J_{(\pm )}{}^i{}_j \na _\pm \phi ^j \,.
\label{eq:jtrans}
\ee
Comparing (\ref{eq:susy}) with (\ref{eq:jtrans}), we can read off $J_{(\pm )}$,
and find that they are both constant, distinct, commuting complex structures.
This is a general feature of complex structures on models constructed
with only chiral and twisted chiral multiplets: the resulting left and right
complex structures must commute \cite{twist}.
In \cite{physlet} it was shown that such commuting structures
exist on \suu , but not on other group manifolds.

A supersymmetric non-linear $\sigma$-model has $N$ left and right handed
supersymmetries when there exist two sets of $N-1$ covariantly
constant complex structures \cite{af,spin}. All the complex structures within
each set anticommute, and the metric has to be
hermitian with respect to all of them. When the
connection has torsion, integrability requires the vanishing of the Nijenhuis
tensors and the left handed (right handed) complex structures have to be
covariantly constant with
respect to the connection consisting of the metric connection plus (minus) the
torsion ($\Gamma _\pm = \{\} \pm T$).

In the case of supersymmetric WZW models, these conditions were completely
solved in \cite{spin}. A complex structure is in one to one correspondence
with a Cartan decomposition of the Lie algebra. On the rootspace, the complex
structure is diagonalized and has eigenvalue $i$ or $-i$, when the root
is positive or  negative, resp.; the Cartan subalgebra is mapped to itself.
The existence of a second complex structure, anticommuting with the first one,
implies a third complex structure (the product of the first two), \ie
$N=3$ implies $N=4$ supersymmetry. It turns
out that $N=4$ is only possible on a restricted set of group manifolds. These
group manifolds are such that they can be written as a product of coset spaces
which have the following structure. Given a group $G$ with Lie algebra $g$
and a Cartan decomposition, we consider the highest root $\theta$. Then
$E_{\pm \theta}$ and $\theta\cdot H$ form a $su(2)$ subalgebra, which we call
$su(2)_\theta$. The remainder of the Cartan subalgebra together with all
roots perpendicular to $\theta$ form another subalgebra $H_\perp$. The coset
space $W=G/H_\perp\times SU(2)_\theta$ is a Wolf space \cite{wolf}. An $N=4$
group manifold can be decomposed as products of coset spaces of the form
$W\times SU(2)_\theta \times U(1)$. The second complex structure
acts within each of these coset spaces. The action on $W$ is clear
as it decomposes in doublets under $SU(2)_\theta$. The action on
$SU(2)\times U(1)$ is such that $E_{\pm\theta}$ get mapped to the
Cartan subalgebra and vice versa.  More details are given in
\cite{spin,proe,st}.

We now analyze the case of \suu\ in detail. Following the discussion above, we
have essentially unique candidates for $J_{(\pm )}$:
\bea
\jpm E_+ = i E_+ \, ,&& \quad \jpm E_- = -i E_- \, ,
\nonu
\jpm (H_0 + i H_3) = \pm i (H_0 + i H_3) \,
 , && \quad
\jpm (H_0 - i H_3) = \mp i (H_0 - i H_3) \, ,
\label{eq:j}
\eea
where $H_0$ generates $U(1)$ transformation and $E_\pm$, $H_3$ are
the generators
of $SU(2)$.  The form is fixed by the condition that $\jp$ and $\jm$ commute.
Eq.\ (\ref{eq:jtrans}) implies analogous relations for the Lie algebra valued
currents:
\be
(g^{-1} \Q _+ g)^a = \jp {}^a{}_b (g^{-1} \na _+ g)^b \, ,
\quad (\Q _- g\, g^{-1} )^a = \jm {}^a{}_b (\na _- g\, g^{-1} )^b \, .
\label{eq:ctrans}
\ee
Using the explicit form of $\jpm$ (\ref{eq:j}), and the relation to the $N=2$
derivatives $D = \half (\na +i \Q )$, $\bD = \half (\na - i \Q )$, we can
lift the relations (\ref{eq:ctrans}) to $N=2$ superspace.
This leads to the following parametrization of $g$ in terms of a chiral
superfield $\Phi$ and a twisted chiral superfield $\Ld$:
\be
g = \frac{e^{i \theta}}{\sqrt{\Phi \bPhi + \Ld \bLd}} \left(
\begin{array}{rc}\Ld \, &\bPhi\\-\Phi \, &\bLd
\end{array} \right) \, ,
\ee
where $\theta = - \half {\rm ln}(\Phi \bPhi + \Ld \bLd )$.
This gives an off-shell $N=2$ formulation of the group \suu . In these
coordinates, the metric on the group manifold is:
\be
ds^2 = \frac{d \Phi d \bPhi + d \Ld d \bLd}{\Phi \bPhi + \Ld \bLd } \, .
\ee
In \cite{twist}, it was shown that the metric can be expressed in terms
of a potential function (analogous to a K\" ahler potential in the case without
torsion): $ds^2 = K_{\Phi \bPhi} d \Phi d\bPhi - K_{\Ld \bLd} d \Ld d\bLd$.
Here, we find
\be
K(\Phi,\bPhi,\Ld,\bLd)= - \int^{\textstyle \frac{\Ld \bLd}{\Phi \bPhi}}
\, \frac{dx}{x}\,{\rm ln}(1+x) +
 \ln \Phi \ln \bPhi \, .
\label{eq:lag}
\ee
This is the $N=2$ superspace lagrangian. We can read off the torsion potential
from $K_{\Phi \bLd}$, etc. (see \cite{twist}).

As noted above, \suu\ actually admits $N=4$ supersymmetry. In $N=2$ superspace,
the necessary condition for $N=4$ supersymmetry is
$ K_{\Phi \bPhi} + K_{\Ld \bLd} = 0$, \cite{twist}, which is clearly
satisfied in this case. In \cite{twist,tom}, the general $N=4$ superspace
description is given. For the case at hand, this was further worked out
in \cite{physlet}.

The existence of a fully off-shell formulation of the model has an important
consequence: it is straightforward to deform the model while maintaining full
$N=4$ supersymmetry, and hence conformal invariance \cite{physlet}.
Such CFT's have recently been proposed as a stringy instanton solutions
\cite{call}.

\vspace{8mm}

We will now take a brief look at the quantum theory and discuss
the $N=4$ superconformal symmetry at the level of on-shell current algebra.

Let us first say a few general things about the on-shell current algebra
in $N=2$ superspace for the supersymmetric WZW model, with level $k$,
on a group $G$ of even dimension. This theory was first worked out
by Hull and Spence in \cite{hs}.

We pick a complex basis for the Lie algebra, labelled by $a$, $\bar{a}$,
$a=1,2,\ldots$, $\half \dim \, G$, which is such that the complex
structure related to the second supersymmetry has eigenvalue
$+i$ on the generators $T_a$ and $-i$ on the generators $T_{\bar{a}}$.
The $N=2$ affine Kac-Moody currents
${\cQ}^a$ and ${\cQ}^{\bar{a}}$ can then be characterized by the following
constraints (we will only discuss the currents that are chiral in the
sense that they are annihilated by $D_-$ and $\bD_-$; for brevity
we will write $D$ for $D_+$ and $\bD$ for $\bD_+$)
\be
D {\cQ}^a=-\frac{1}{2(k+\Ht)} {f^a}_{bc} {\cQ}^b {\cQ}^c, \quad
\bD {\cQ}^{\bar{a}}=-\frac{1}{2(k+\Ht)} {f^{\bar{a}}}_{\bar{b}
     \bar{c}} {\cQ}^{\bar{b}} {\cQ}^{\bar{c}} \,.
\label{eq:cons}
\ee
In here $\Ht$ is the dual Coxeter number of $G$ and the the $f$'s
are the structure constants in the complex basis.
The fundamental OPE's of these $N=2$ superfields are
\bea
 {\cQ}^a (Z_{1}) {\cQ}^b (Z_{2})   & = & \tzb {f^{ab}}_{c} {\cQ}^c
 (Z_{2})+\tzbb \frac{1}{k+\Ht} {f^{a}}_{ec} {f^{be}}_{d}
 {\cQ}^c {\cQ}^d (Z_{2})
\nonu
 {\cQ}^{\bar{a}} (Z_{1}) {\cQ}^{\bar{b}} (Z_{2})  & = &
 \tz {f^{\bar{a} \bar{b}}}_{\bar{c}} {\cQ}^{\bar{c}} (Z_{2})
 -\tzbb \frac{1}{k+\Ht} {f^{\bar{a}}}_{\bar{e} \bar{c}}
 {f^{\bar{b} \bar{e}}}_{\bar{d}} {\cQ}^{\bar{c}} {\cQ}^{\bar{d}} (Z_{2})
\nonu
 {\cQ}^{a} (Z_{1}) {\cQ}^{\bar{b}} (Z_{2})  & = & (k+\Ht)
 \left[ \frac{g^{a \bar{b}}}{z_{12}} + \tzzbb (\frac{1}{2} g^{a \bar{b}}
        + \frac{1}{2(k+\Ht)} {f^a}_{cd} f^{\bar{b} cd}) \right]
\nonu
 & & + \tz {f^{a \bar{b}}}_c {\cQ}^c (Z_{2}) + \tzb
 {f^{a \bar{b}}}_{\bar{c}} {\cQ}^{\bar{c}} (Z_{2})
\nonu
 & & + \tzbb \left[ {f^{a \bar{b}}}_c \bar{D} {\cQ}^c (Z_{2})
 + \frac{1}{k+\Ht} {f^{a \bar{c}}}_{d} {f^{\bar{b}}}_{\bar{c} \bar{e}}
 {\cQ}^d {\cQ}^{\bar{e}} (Z_{2}) \right],
\eea
where
\be
\th_{12}=\th_{1}-\th_{2}, \; \bth_{12}=\bth_{1}-\bth_{2}, \;
\mbox{and} \;
z_{12}=z_{1}-z_{2}-\frac{1}{2}(\th_{1} \bth_{2} + \bth_{1} \th_{2}).
\ee

Let us now focus on the $N=2$ superconformal algebra.
The appropriate generalization to $N=2$ superspace of the well-known
Sugawara construction gives the following formula for the $N=2$ super
stress tensor in terms of the super Kac-Moody currents ${\cQ}^a$ and
${\cQ}^{\bar{a}}$ (\cite{hs})
\be
{\cal T}= \frac{i}{k+\Ht} g_{a \bar{b}}( {\cQ}^a {\cQ}^{\bar{b}} )
  -\frac{1}{k+\Ht} ( f_{\bar{a}} D {\cQ}^{\bar{a}} + f_a \bD {\cQ}^a) \,,
\ee
where $g_{a \bar{b}}=\delta_{a \bar{b}}$,
$f_a = g_{b \bar{c}} f_a{}^{b \bar{c}}$ and
$f_{\bar{a}}= g_{b \bar{c}} f_{\bar{a}}{}^{b \bar{c}}$.
It satisfies the OPE
\be
{\cT} (Z_{1}) {\cT} (Z_{2})= - \frac{1}{z^{2}_{12}} \frac{c_k}{3}
  -i \left[ \tzzbb +\tz D-\tzb \bD +\tzbb \partial_2 \right] {\cT}(Z_{2}) \,.
\label{eq:TT}
\ee
The total central charge is the sum of contributions
$c^\prime_k = \frac{3 \,\dim G}{2} (1-\frac{2 \Ht}{3 (k+\Ht)})$ for each
simple factor of $G$. Of course, the $N=2$ superfield ${\cal T}$ has
as its component fields the bosonic stress tensor $T$, two supercurrents
$G$ and $\overline{G}$ and the $U(1)$ current $J$, which together form
the familiar $N=2$ current algebra.

In the example of $G=SU(2)\times U(1)$, the affine Kac-Moody
currents ${\cQ}^a$ and ${\cQ}^{\bar{a}}$ can be expressed in the
coordinate fields $\Ld$, $\bLd$, $\Phi$, and $\bPhi$ as follows
\bea
  {\cQ}^1   & =& \frac{(k+2)}{r^2}({\Phi} D {\Ld}-{\Ld} D {\Phi})
\nonu
  {\cQ}^2   & =& -\frac{i(k+2)}{r^2}({\bLd} D {\Ld}+{\bPhi} D {\Phi})
\nonu
  {\cQ}^{\bar{1}} & =& -\frac{(k+2)}{r^2}({\bPhi} \bD {\bLd}-{\bLd} \bD
  {\bPhi})
\nonu
  {\cQ}^{\bar{2}} & =& -\frac{i(k+2)}{r^2}({\Ld} \bD {\bLd}+{\Phi} \bD
  {\bPhi}) \,,
\label{eq:Qss}
\eea
where
\be
  r^2={\Ld} {\bLd}+{\Phi} {\bPhi} \,.
\ee
Via the above they lead to an $N=2$ superconformal algebra of
central charge $c_k = \frac{9}{2}(1-\frac{4}{3(k+2)})+\frac{3}{2}
=6 \, \frac{k+1}{k+2}$,
which is $c=4$ for $k=1$ and approaches $c=6$ if $k \rightarrow \infty$.

The above makes manifest the $N=2$ superconformal symmetry of our
model. However, we already mentioned that the model actually
posseses a $N=4$ superconformal symmetry. The appropriate algebra
is the so-called `large' $N=4$ superconformal algebra
\cite{ad}-\cite{stp}. This algebra
has 16 generators, which are: the spin-2 stress tensor $T$,
4 spin-3/2 supercurrents $G^i$, 7 spin-1 currents generating
the affine extension of $SU(2)\times SU(2) \times U(1)$ and 4 spin-1/2 currents
$\Gamma^i$. The unitary representations of this algebra can be
characterized by two integers $k_+$ and $k_-$, with a corresponding
central charge equal to $c(k_+,k_-) = 6 \, k_+ k_-/(k_+ + k_-)$.
The parameter $\alpha= \half \frac{k_+ - k_-}{k_+ + k_-}$
is a measure for the asymmetry between two affine $SU(2)$ subalgebras,
which have level $k_+$ and $k_-$, respectively. The projective subalgebra
is isomorphic to $D(2,1; \alpha-\half)$.

It was shown in \cite{stp,stps} that the level $k$ \suu\ WZW model gives
a realization of this $N=4$ superconformal algebra with
$k_+=(k+1)$, $k_-=1$. (For $k=0$ the bosonic $SU(2)$ WZW model
decouples and this realization reduces to the $c=3$ realization
with one free boson and four free fermions which was first
discussed in \cite{sch1}). We will now derive explicit formulas for
the generators of the full $N=4$ algebra in terms of the fundamental
superfields $\Ld$ and $\Phi$ of the model.

When written in (chiral) $N=2$ superspace, the full $N=4$ algebra
is generated by (i) the super stress tensor $\cT$, which
has conformal spin 1, (ii) two spin-1/2 superfields $\cA$ and
$\cB$ and (iii) a spin-0 superfield $\cI$. Each of these provides
four component fields, so that we find the correct total number
of 16 currents.

To determine the extra currents $\cA$, $\cB$ and $\cI$,
we will use the results of \cite{st}, where the relation between
the affine currents and the $N=4$ superconformal algebra was
worked out in detail in $N=1$ superspace.
The explicit relation between the $N=1$ super Kac-Moody currents
$Q^a$, $Q^{\bar{a}}$ in chiral $N=1$ superspace $(z,\theta^1)$
\cite{paolo} and the $N=2$ super Kac-Moody currents is as follows
(there are actually equal numbers of both since the $N=2$
affine currents are constrained, see (\ref{eq:cons})), \cite{hs}
\bea
 {\cQ}^a  & = & Q^a-i {\theta^2} (\nabla Q^a+\frac{1}{k+\Ht}
                {f^a}_{bc}Q^b Q^c)
\nonu
{\cQ}^{\bar{a}} & = & Q^{\bar{a}}+i {\theta^2} (\nabla Q^{\bar{a}}+
        \frac{1}{k+\Ht} {f^{\bar{a}}}_{\bar{b} \bar{c}}
        Q^{\bar{b}} Q^{\bar{c}}).
\label{eq:Qs}
\eea
In here, $\theta^2$ is the second fermionic coordinate.
These relations together with the results in \cite{st} make it possible
to determine the extra currents. We find
\bea
D {\cI} = \frac{1}{\sqrt{1-4\alpha^2}} \frac{i}{k+2}{\cQ}^{2},
&& \quad
\bD {\cI} = \frac{1}{\sqrt{1-4\alpha^2}}\frac{i}{k+2} {\cQ}^{\bar{2}}
\nonu
{\cA} = -\frac{1}{2}({\cQ}^1-{\cQ}^{\bar{1}}),
&& \quad
{\cB} = \frac{i}{2}({\cQ}^1+{\cQ}^{\bar{1}}) \,,
\eea
where $\alpha=\frac{k}{2(k+2)}$. Together with the expression
for the stress energy tensor,
\be
{\cal T} = \frac{i}{k+2}
  ( {\cQ}^1 {\cQ}^{\bar{1}}+{\cQ}^2 {\cQ}^{\bar{2}})
  - \frac{1}{k+2} (D{\cQ}^{\bar{2}}-\bD {\cQ}^{2}) \,,
\ee
these relations express all the generators of the $N=4$ superconformal
algebra in terms of the $N=2$ affine Kac-Moody currents and thereby,
through (\ref{eq:Qss}), in terms of the fundamental fields.
Comparing with (\ref{eq:Qss}),
we may conclude that ${\cI}$ is given by
\be
{\cI} = \frac{1}{\sqrt{1-4\alpha^2}} \ln r^2 \,.
\ee
The OPE's of the currents ${\cT}$, ${\cA}$, ${\cB}$ and ${\cI}$
are given by (\ref{eq:TT}) and
\bea
{\cT} (Z_{1}) {\cI} (Z_{2}) &=&
  -i \left[ \tz D - \tzb \bD +\tzbb \partial_2 \right]
 {\cI} (Z_{2})
\nonu
{\cT} (Z_{1}) {\cA} (Z_{2}) &=&
  -i \left[ \tzzbb \frac{1}{2} +\tz D -\tzb \bD
            + \tzbb \partial_2 \right] {\cA} (Z_{2})
  - \frac{2\alpha}{z_{12}} {\cB} (Z_{2})
\nonu
{\cT} (Z_{1}) {\cB} (Z_{2}) &=&
  -i \left[ \tzzbb \frac{1}{2} +\tz D -\tzb \bD
            + \tzbb \partial_2 \right] {\cB} (Z_{2})
  + \frac{2\alpha}{z_{12}} {\cA} (Z_{2})
\nonu
{\cA} (Z_{1}) {\cA} (Z_{2}) &=&
  - \frac{c_1}{3} \, \frac{1}{z_{12}} - \tzbb \frac{i}{2}
  \left[ {\cT}(Z_{2})+\frac{2\alpha i}{\sqrt{1-4\alpha^2}}
  [D,\bD]{\cI}(Z_{2}) \right]
\nonu
{\cB} (Z_{1}) {\cB} (Z_{2}) &=&
  - \frac{c_1}{3} \, \frac{1}{z_{12}} -\tzbb \frac{i}{2}
  \left[ {\cT}(Z_{2})+\frac{2\alpha i}{\sqrt{1-4\alpha^2}}
  [D,\bD]{\cI}(Z_{2}) \right]
\nonu
{\cA} (Z_{1}) {\cB} (Z_{2}) &=&
  - \frac{c_2}{3} \, \tzzbb \frac{i}{2}
\nonu
  && -i \left[ \tz D - \tzb \bD + \frac{1}{2} \tzbb \partial_{2} \right]
     \frac{1}{\sqrt{1-4\alpha^2}} {\cI} (Z_{2})
\nonu
{\cI} (Z_{1}) {\cA} (Z_{2}) &=&
  \tzbb \frac{i}{2} \sqrt{1-4\alpha^2} \, {\cB} (Z_{2})
\nonu
{\cI} (Z_{1}) {\cB} (Z_{2}) &=&
  - \tzbb \frac{i}{2} \sqrt{1-4\alpha^2} \, {\cA} (Z_{2})
\nonu
{\cI} (Z_{1}) {\cI} (Z_{2}) &=&
  - \frac{c_k}{3} \, \ln z_{12} \,,
\eea
where $c_1 = \frac{1}{1-4 \alpha^2} c_k = \frac{3}{2} \, (k+2)$ and
$c_2 = \frac{2 \alpha}{1-4 \alpha^2} c_k = \frac{3}{2} \, k$.

\vspace{8mm}
We now return to the off-shell $N=2$ superspace action given in (\ref{eq:lag}).
This action has the form that admits a duality
transformation \cite{shout,twist,toms}.  From the general theory, we know
that after the transformation, all the superfields will be chiral, and the
manifold will therefore be K\"ahler (with vanishing torsion). We will now
explicitly compute the metric of this manifold.

The first step of the duality transformation is to rewrite the
action (\ref{eq:lag}) in a first order form. The first order lagrangian
depends on the chiral superfields $\Phi$, $\bPhi$ and  $\eta$ and $\bareta$
and on the real quantity $X$. We define
\bea
\tK_X (\Phi,\bPhi,\eta,\bareta)
&=& - \int^{e^X}
\, \frac{dx}{x}\,{\rm ln}(1+x) +
 \ln \Phi \ln \bPhi \nonu
  && + \, a \, [X +   \ln (\Phi \bPhi)] (\eta + \bareta) \,,
\label{eq:lag1}
\eea
where $a$ is a constant $\neq 0$.
When varying the first order action w.r.t.\ $\eta$ and $\bareta$,
we should keep in mind that these are {\it constrained}\ superfields.
It can be shown that the most general expression for $X$ that is compatible
with the $\eta$, $\bareta$ field equations is
\be
X = \ln (\Ld\bLd )-\ln (\Phi\bPhi ) \,,
\ee
where $\Ld$, $\bLd$ is a twisted chiral superfield. Substituting this
back into the first-order action (\ref{eq:lag1})
one finds back the original action (\ref{eq:lag}).

Let us now treat the first order action differently, and use the
field equations of the field $X$ instead of those of $\eta$, $\bareta$.
They lead to
\be
 1 + e^X = e^{a\,(\eta+\bareta)} \,.
\ee
We now define the following variables
\be
\chi = e^{a\, \eta}, \quad
\tPhi = \ln \Phi + a \eta \,.
\ee
Notice that both $\chi$ and $\tPhi$ are {\it chiral} $N=2$ superfields.
Substituting the above into the first order action (\ref{eq:lag1}),
we arrive at the following second order action
\bea
\tK(\tPhi,\tbPhi,\chi,\bchi) &=&
  - \int^{\chi \bchi -1} \frac{dx}{x} \ln (1+x)
  + \ln (\chi\bchi-1) \ln (\chi \bchi)
\nonu
  && - \half (\ln (\chi\bchi ))^2\, +\, \tPhi \tbPhi \,.
\label{eq:lag2}
\eea
This lagrangian describes a theory which is dual to the original
theory. Although both theories are equivalent at the level of the
classical equations of motion, their geometric interpretation is
very much different: the original WZW model describes a group manifold
(with torsion), whereas the dual model describes a K\"ahler geometry
(without torsion) with the K\"ahler potential given by (\ref{eq:lag2}).

The geometry associated with the dual model clearly splits as a product
of a torus (with coordinates $\tPhi$ and $\tbPhi$) and
a disk bearing the singular metric
\be
ds^2 = \tK_{\chi\bchi} \, d\chi \, d\bchi
     =  \frac{d\chi d\bchi}{\chi\bchi (\chi\bchi -1)}  \, .
\ee
In terms of the coordinates $u=1/\chi$ this metric takes the simple form
\be
ds^2 = \frac{du \, d\overline u}{1-u\overline u} \, .
\label{eq:metric}
\ee
If one follows the process of passing from the original to the dual
formulation at the level of the functional integral, one finds that,
apart from the change of metric, the transition leads to a non-vanishing
dilaton field in the dual formulation (see \cite{toms} for a careful
discussion). In our case the dilaton field is given by
\be
\phi = \ln(1-u\overline u) \,.
\ee

It can be observed that the above combination of metric and dilaton
fields is such that the sigma model is conformally invariant.
Due to this, this geometry can serve as a consistent background
for a string theory with a two-dimensional target space-time.
This observation has been worked out by Witten \cite{black}, who
proposed the interpretation of this geometry as a back hole solution to
$D=2$ string theory. We would like to remark that our derivation
of this geometry (through a duality transformation in $N=2$ superspace)
is similar to, although independent from, Witten's derivation, which
is based on a gauging a $U(1)$ subgroup in the $SU(2)$ WZW model.
(The relation has recently been clarified in \cite{dqc}.)

It would be interesting to work out the duality transformation
at the quantum level. The central charge $c_k$ can be written
as $c_k = 3 + \frac{3k}{k+2}$, where the $c=3$ part corresponds
to the free fields $\tPhi$, $\tbPhi$ and the remaining part
describes the interacting sigma-model with metric (\ref{eq:metric}).
The fate of the $N=4$ superconformal
symmetry in the dual model is not yet clear. On first inspection, one finds
that the duality breaks the $N=4$ supersymmetry, but one still expects
that some remnant of it could survive and might have some interesting
applications in the (super)string theory interpretation of the model.
We leave these issues for further study.

\end{document}